\documentclass[11pt,a4paper]{article}
\usepackage{jcappub}
\usepackage{amsmath}
\usepackage{amsfonts}
\usepackage{amssymb}
\usepackage[utf8]{inputenc}
\usepackage{graphicx}
\usepackage{multirow}
\allowdisplaybreaks
\title{Ultraviolet Unitarity Violations in Non-minimally Coupled Scalar-Starobinsky Inflation}

\author{Sukanta Panda,}
\author{Abbas Altafhussain Tinwala,}
\author{and Archit Vidyarthi}

\affiliation{Indian Institute of Science Education and Research Bhopal - 462066, India}

\emailAdd{sukanta@iiserb.ac.in}
\emailAdd{abbas18@iiserb.ac.in}
\emailAdd{archit17@iiserb.ac.in}

\abstract{We perform the calculation for tree-level ultraviolet unitarity violation scales for scalar-$R^2$ inflation models by including an additional $R^2 |\Phi|^2$-type term. Due to certain constraints, we resort to the Einstein frame for our calculations, where we separate our analysis between metric and Palatini formulations. We follow recent works in this line that debunk the naive predictions for unitarity violations in Higgs' inflation models to determine how to accurately estimate the behaviour of scattering amplitudes in the UV limit. Later, we work out different cases by assuming potentials corresponding to known inflation scenarios so we could predict the range of coupling parameters for which the theories would remain unitary up to the Planckian regime. We also try to find the behaviour of the scattering amplitudes for these theories during the transition from inflationary to reheating epoch.}

\arxivnumber{}

\begin{document}
\maketitle
\flushbottom
    \section{Introduction}
    Inflationary paradigm \cite{Brandenberger:1992dw,Brandenberger:1992qj,Brandenberger:1993zc,Liddle:2000cg,riotto2002inflation,linde2005particle,Mukhanov:2005sc,Sriramkumar2009AnIT,Sasaki:2012ss,sato_inflationary_2015,Maroto:2015dqa} has withstood the test of time when it comes to successfully explaining the current state of the universe to a large extent. But even with that kind of accuracy, a deeper understanding of the mechanism of inflation still eludes our best efforts. The Standard Model of cosmology assumes a canonical scalar field that serves as the inflaton and drives inflation. And while this simple model has been able to explain a variety of problems at both background and quantum levels, there are still subtleties that remain.
	
	To explain away the remainder of the issues, multitudes of inflationary models have been proposed over the past few decades \cite{Capozziello:2011et,Nojiri:2017ncd}. For modifications on the matter side, these include noncanonical scalar models \cite{Ramirez2019NoncanonicalAT}, or models that replace the scalar field with, say, a massive vector field \cite{Panda:2021cbf,DeFelice:2016yws,Heisenberg:2016wtr,Nakamura:2017dnf}, fermionic field \cite{deSouza:2008az,Grams:2014woa,Benisty:2019pxb,Shokri:2021aum}, antisymmetric tensor field \cite{Aashish:2018lhv,Aashish:2019zsy,Aashish:2020mlw,Aashish:2021btg,Aashish:2021gdf}, and models involving nonminimal couplings with these fields, all of which require the introduction of new physics. Apart from these, there have been several propositions that hold the Standard Model of particle physics to be fundamental and intend not to modify that and instead propose modifications on the gravity side. These modifications may include the inclusion of higher derivative terms such as Starobinsky model \cite{Starobinsky:1980te,Kaneda:2015jma,Calmet:2016fsr,Pi2017ScalaronFR,Gundhi:2018wyz,Gialamas:2019nly} or Horndeski theories \cite{Horndeski:1974wa,Kobayashi:2019hrl,Joshi:2021azw}, and Gauss-Bonnet gravity \cite{PhysRevLett.124.081301,Fernandes:2022zrq}, or nonlocal terms \cite{Maggiore:2016gpx,Belgacem:2018ypx,Joshi:2019cyk,Joshi:2022dpz} which attempt to get rid of issues that plague Einstein-Hilbert gravity. There are also models that incorporate both these features simultaneously in order to fix the problems encountered in one model with features from another, for example the Scalaron-Higgs models \cite{Kaneda:2015jma,Salvio:2015kka,Salvio:2016vxi,EMA2017403,PhysRevD.96.123506,Gundhi:2018wyz,He_2018,Canko:2019mud,Ema_2020,Ema_2021,He_2021}, which shall be the topic of interest in this work.
	
	Keeping matters simple, the Standard Model of particle physics offers only one scalar field that could be used as a candidate for inflaton: the Higgs' field (without resorting to new physics) \cite{Lerner:2009na,Lerner:2010mq,Atkins:2010yg,Germani:2010gm,Calmet:2013hia,Calmet:2016fsr,Rubio:2018ogq,Hamada:2020kuy,McDonald_2021}. Apart from that, Starobinsky model involves transforming the $R^2$-term to a scalar field (called scalaron) which provide the same kind of dynamics as scalar-driven inflation models. These two are currently the best candidates for explaining the inflationary dynamics without introducing new fields into the picture. Interestingly, they have been previously shown to be equivalent even at one-loop level \cite{Ketov:2019toi,Ketov:2022qwj} (a general on-shell equivalence of $F(R)$ gravity models and scalar-tensor theories was proved in \cite{Ruf:2017xon} and further in both Jordan and Einstein frames in \cite{Ohta_2018}).
	
	Possible tree-level unitarity violations posed by Higgs' field inflation models have been subject to debate for decades now \cite{Lerner:2010mq,Calmet:2013hia,Escriva:2016cwl,Rubio:2018ogq,McDonald_2021}. The amount of interest that both Higgs' and Starobinsky models respectively garner is because of their agreement with the observed data available to us. But the question of their tree-level unitarity has persisted all the while. Recently, it was shown in \cite{McDonald_2021} that the dimension counting approach that we normally employ to figure out possible unitarity violation scales might be flawed. They showed explicitly by working with a complex Higgs' singlet in Palatini formulation that all possible unitarity violations are curbed as we approach the naive scales, implying that the theory might be safe up to the Planck energy scale. We will briefly review their work in Sec. \ref{unitarity} before moving on to the main text.
	
	While the paper doesn't establish their method as being foolproof, they do suggest that the current techniques are flawed and need to be revised to get a better idea of the state of unitarity violations. This work was followed up by \cite{Antoniadis2021} where the authors performed a similar analysis for Higgs' inflation models in both metric and Palatini formulations, also including the Starobinsky term in the models. More recently, \cite{antoniadis2022addendum} was produced as an addendum to \cite{Antoniadis2021} where the authors generalized their findings to pose better arguments against the validity of \cite{McDonald_2021}.
	
    In this work, we shall be using a similar method to find unitarity violation scales for a general class scalar-tensor inflation theories. The coupling at the center of our analysis in this model was proposed in \cite{Mathew:2017lvh} where the authors were able to narrow down the constraints on the theory to ensure a safe exit from the inflationary epoch. It was followed up by \cite{Das2021} where the authors assumed a variety of potentials and couplings and determined the viability of each in the Palatini formulation. The action in (\ref{action}) was concluded to be an attractive possibility. By analyzing the tree-level unitarity of the theory (\ref{action}), we intend to further solidify the previous claims and to check whether unitarity constraints are further able to restrict the range of the parameters involved.
	
	The paper is laid out as follows: first, we give a brief overview of the current state of the unitarity violation issue that pertains to Higg's inflation models in Section \ref{unitarity}; then, in Section \ref{secmodel} we describe the model that we'll be working with and motivate the reason for moving to the Einstein frame; since this is where the metric and Palatini formulations diverge and have been worked out separately in Sections \ref{metric} and \ref{pala}, respectively; finally, we briefly recapitulate the work, draw conclusions from the results obtained, and make some closing statements.

    \section{Unitarity Violations?}\label{unitarity}
    Naively speaking, it appears that the unitarity violation scales are easy to identify for all cases considered in this paper (for example, using dimension counting methods). However, the argument put forward by the author in \cite{McDonald_2021} about the breakdown of perturbative expansion in the high-energy limit implies that naive guesses may point to an incomplete picture. In this section, we first lay down the basics and assumptions involved in their approach, before applying them to our model(s) to gain some valuable insight into the issue of unitarity violations.
    
    A scattering amplitude gains energy scaling from the presence of kinetic interaction terms present in the action. In $s$-, $t$-, $u$-channel diagrams, even though the propagators of canonicalized fields are obtained using kinetic terms, the vertex factors from derivative couplings ensure that the energy scaling is always polynomial (at tree-level).
    
    The author in \cite{McDonald_2021} considers a 2 scalar $\to$ 2 scalar scattering process where the particles are assumed to be at an energy $\sim E$ (but not exactly $E$). For such a case, the incoming particles would be described as wavepackets with energy spread $\Delta E\sim E$. Conversely, $\Delta x\sim 1/\Delta E\sim 1/E$, owing to uncertainty. Then, the vacuum expectation value of the energy density becomes: $\left<\rho\right>\sim E/(\Delta x)^3\sim E^4$. From field theory, we know that in the interaction volume $(\Delta x)^3$, $\left<\rho\right>\sim E^2\left<\phi^2\right>$, i.e. the vacuum expectation value of field $\left<\phi\right>\sim E$, where $\phi$ is one of the external scalar fields involved in the scattering process. The author's work is centered around the following assumptions: that the scattering amplitude is unaffected by the introduction of a background field, and that the form of the scattering amplitude is preserved even when the background field scale approaches the energy of dynamical fields.
    
    For reasons that we later lay out in Section \ref{secmodel}, we stick with the Einstein frame for our calculations. It has been explicitly shown for the case of a Higgs' inflation model that the final scattering amplitude remains frame invariant \cite{Antoniadis2021}. However, in \cite{McDonald_2021}, the author lays out the reason to only consider unitarity violation scales in the Jordan frame. They claim that in order to move to the Einstein frame, we use a Weyl rescaling function $\Omega^2$ where we assign the dynamical fields a magnitude $E$ that could assume any scale relative to the field background $\Bar{\phi}$ to obtain a perturbative expansion. Since the Jordan frame is devoid of any such requirement, this implies that we have more freedom to adjust the energy scale of the dynamical fields in the Jordan frame as compared to the Einstein frame. Using the kinetic terms of scattered fields, before expanding around a background, the energies in the Einstein frame ($E_E$) are related to those in the Jordan frame ($E_J$) by the expression $E_E=\frac{E_J}{\Bar{\Omega}}$, where $\Bar{\Omega}$ is background part in the perturbative expansion of $\Omega$. 
    
    The author in \cite{McDonald_2021} proposed that we work around a large inflaton background $\Bar{\phi}\gg\frac{M_P}{\sqrt{\xi}}$. This allowed them to expand $\Omega^{-1}$ and $\Omega^{-2}$ perturbatively as obtained in their work as well as in \cite{Antoniadis2021}. However, this expansion can only be made when the expectation value of the dynamical fields $\left<\phi\right>\sim E$ is small compared to the background $\Bar{\phi}$. Later, to get the full picture of unitarity violations, the author in \cite{McDonald_2021} suggested that we work with the assumption that the amplitude that we've obtained works just as well for the condition $E\gg \Bar{\phi}$, provided it's a continuous function, and then proceed to take the limit $\Bar{\phi}\to E$. This procedure will be referred to as `matching' for the rest of the paper. It should be abundantly clear that the matching procedure is only applicable when the ranges of $E$ and $\Bar{\phi}$ overlap as we approach the Planck scale.
    
    This approach was later modified to generalize the condition on the inflaton background in the recently published addendum \cite{antoniadis2022addendum}. For this, while we do still perform a perturbative Taylor expansion, it is no longer truncated to include the relevant number of terms. Instead, we sum the full infinite series of terms to obtain non-trivial form factors as coefficients to the relevant terms. This method has the flexibility that it allows for us to  obtain results around a small background as well, which is a useful limit since it allows us to paint a picture for the reheating era of the universe where $\Bar{\phi}\to0$. Also, it allows us to circumvent the assumption made in \cite{McDonald_2021} regarding the continuity of the scattering amplitude across the regimes $E\ll\Bar{\phi}$ and $E\gg\Bar{\phi}$. Next, we shall look at the model under consideration.

    \section{Model: Why go to the Einstein Frame?}\label{secmodel}
    The model that we consider for the purpose of this paper is:
    \begin{equation}\label{action}
	   S = \int d^4x \sqrt{g} \left[\dfrac{M_P^2}{2}\left(1+\frac{\xi|\Phi|^2}{M_P^2}\right)\left(R+\frac{\alpha}{2M_P^2}R^2\right)-|\partial \Phi|^2\right],
	\end{equation}
	where $\Phi=\frac{1}{\sqrt{2}}(\phi_1+i\phi_2)$ represents the complex scalar field. We assume no potential or mass term for the scalar like the authors in \cite{Antoniadis2021, McDonald_2021} since we are only interested in unitarity violations stemming from the non-minimal coupling, which is also used to defined the naive unitarity violation scale for the model. Later, we shall assume different potentials, as examples, which would impose certain constraints on the coupling parameters $\xi$ and $\alpha$ used in the theory, for which we can refer to the results obtained using the analysis for a general case.
	
	The reason for working with a complex scalar is because it has been shown previously in the work of \cite{Hertzberg:2010dc} that theories involving a single scalar field coupled with gravity are well-behaved up to the Planckian regime despite the fact that they may naively exhibit violations. Taking cues from the calculation performed by the authors in \cite{Antoniadis2021}, we shall be working in the weak gravity limit: $g_{\mu\nu}=\eta_{\mu\nu}+\kappa h_{\mu\nu}$; and expand one of the scalar fields around a large background value $\phi_1 = \Bar{\phi}_1+\varphi_1$, where $\Bar{\phi}_1$ represents the classical background, and $\varphi_1$ represents the quantum fluctuations around it. 
	
	The paper \cite{McDonald_2021} used a large inflaton background governed by the condition $\Bar{\phi}_1\gg\frac{M_P}{\sqrt{\xi}}$. We'll be employing a general background in this paper, following the analysis in \cite{antoniadis2022addendum}, and also look at the $\Bar{\phi}\ll\frac{M_P}{\sqrt{\xi}}$ limit simultaneously. We won't impose any constraints on the parameters $\xi$ and $\alpha$ and, therefore, expect to find some constraints based on unitarity arguments as part of this analysis.
	
	Expanding (\ref{action}) up to second order,
	\begin{align}\label{1order}
	    \mathcal{L}^{(1)}=&\frac{\kappa M_P^2}{2}\left(1 +\frac{\xi\Bar{\phi}_1^2}{M_P^2}\right)(\partial_\mu\partial_\nu h^{\mu\nu}-\Box h),\\
	    \mathcal{L}^{(2)}=&-\frac{\kappa^2M_P^2}{8}\left(1 +\frac{\xi\Bar{\phi}_1^2}{M_P^2}\right)(\partial_\rho h^{\mu\nu}\partial^\rho h_{\mu\nu} +\partial_\mu h\partial^\mu h-2\partial_\mu h \partial_\nu h^{\mu\nu} +2\partial^\mu h^{\nu\rho}\partial_\nu h_{\mu\rho})\nonumber\\&
	    +\frac{\alpha\kappa^2}{4}\left(1 +\frac{\xi\Bar{\phi}_1^2}{M_P^2}\right)\left(2\Box h \Box h -4\partial_\mu\partial_\nu h^{\mu\nu}\Box h+2\partial_\mu\partial_\nu h^{\mu\nu}\partial_\rho\partial_\sigma h^{\rho\sigma}\right)\nonumber\\&
	    +\kappa\xi\Bar{\phi}_1\varphi_1(\partial_\mu\partial_\nu h^{\mu\nu}-\Box h)-\frac{1}{2}\partial_\mu\varphi_1\partial^\mu\varphi_1.\label{2order}
	\end{align}
	The first term of (\ref{2order}) gives the kinetic term for gravitons as we set $\kappa^2=4\left[M_P^2\left(1+\frac{\xi\Bar{\phi}_1^2}{M_P^2}\right)\right]^{-1}$. The second term in (\ref{2order}) contains contributions from the Starobinsky correction, characterized by the higher order derivative operators and the presence of $\alpha$ in their coefficients. The second last term contains kinetic mixing between scalar and graviton perturbations. Setting $\alpha= 0$, the kinetic matrix obtained for (\ref{2order}) can be easily diagonalized and the corresponding field transformations are well established in the literature \cite{Rubio:2018ogq,Antoniadis2021}. Otherwise, however, it can be explicitly shown that the system cannot be diagonalized, at least in a meaningful way that would benefit the rest of the calculation. Had we considered a scalar coupling with, say, the Gauss-Bonnet term, due to the absence of higher derivative terms at second order perturbative expansion, we would expect the diagonalization to be straightforward.
	
	As such, we instead resort to the method that involves introducing an auxiliary scalar field in the action to take care of the higher order curvature terms. Rewriting the action (\ref{action}), we get,
	\begin{align}\label{aux}
	    S=\int d^4x \sqrt{-g} \left[\frac{1}{2}M_P^2 f(R,\phi_1,\phi_2)-\frac{1}{2}\partial_\mu\phi_1\partial^\mu\phi_1-\frac{1}{2}\partial_\mu\phi_2\partial^\mu\phi_2\right],
	\end{align}
	where,
	\begin{equation}\label{fg}
	    f(R,\phi)=\left(1+\frac{2\xi|\Phi|^2}{M_P^2}\right)\left(R+\frac{\alpha}{2M_P^2}R^2\right)=G(\phi_1,\phi_2)\left(R+\frac{\alpha}{2M_P^2}R^2\right).
	\end{equation}
	Introducing an auxiliary scalar into the action,
	\begin{align}\label{lagrange}
	    S=&\int d^4x \sqrt{-g} \left[\frac{1}{2}M_P^2 f(\chi,\phi_1,\phi_2) +\frac{1}{2}M_P^2\frac{df}{d\chi^2}(R-\chi^2)-\frac{1}{2}\partial_\mu\phi_1\partial^\mu\phi_1-\frac{1}{2}\partial_\mu\phi_2\partial^\mu\phi_2\right]\nonumber\\
	    =&\int d^4x \sqrt{-g} \left[\frac{1}{2}M_P^2 f'(\chi,\phi_1,\phi_2) R -W(\chi,\phi_1,\phi_2)-\frac{1}{2}\partial_\mu\phi_1\partial^\mu\phi_1-\frac{1}{2}\partial_\mu\phi_2\partial^\mu\phi_2\right],
	\end{align}
	where $f'(\chi,\phi_1,\phi_2)=\frac{df}{d\chi^2}$ and,
	\begin{equation}
	    W(\chi,\phi_1,\phi_2)=\frac{1}{2}M_P^2\left[\chi^2f'(\chi,\phi_1,\phi_2)-f(\chi,\phi_1,\phi_2)\right]=\frac{\alpha}{4}G(\phi_1,\phi_2)\chi^4.
	\end{equation}
	At this stage, since the higher derivative terms from the gravity side have been `transformed' into scalar degrees of freedom, we could expect a more favorable outcome when performing a perturbative analysis. However, all the $\alpha$ dependence lies in $W(\chi,\phi_1,\phi_2)$ which contains 5- and 6-vertices, of which four of the external arms belong to $\chi$, which is a non-propagating field in (\ref{lagrange}). Therefore, it is easy to see that all relevant information that we need from the additional Starobinsky term and its interaction with $G(\phi_1,\phi_1)$ is lost, at least at the level of our computation. Ruling out the Jordan frame for the aforementioned reasons, we now perform the following Weyl rescaling to move to the Einstein frame:
	\begin{equation}\label{weyl}
	    g_{\mu\nu}\to f'(\chi,\phi_1,\phi_2)g_{\mu\nu}.
	\end{equation}
    This is where the results of the metric and Palatini formulation diverge, since the Ricci scalar transforms differently both. Both cases have been worked out in detail in separate sections.

	\section{Metric Formulation}\label{metric}
    We begin with the metric calculation first. For this case, the Weyl transformed action becomes:
    \begin{multline}\label{einmetric}
	    S=\int d^4x \sqrt{-g} \left[\frac{1}{2}M_P^2R-\frac{3}{4}\frac{M_P^2}{f'^2(\chi,\phi_1,\phi_2)}\partial_\mu f'(\chi,\phi_1,\phi_2)\partial^\mu f'(\chi,\phi_1,\phi_2)\right.\\\left.-\frac{1}{2}\left(\frac{\partial_\mu\phi_1\partial^\mu\phi_1}{f'(\chi,\phi_1,\phi_2)}\right)-\frac{1}{2}\left(\frac{\partial_\mu\phi_2\partial^\mu\phi_2}{f'(\chi,\phi_1,\phi_2)}\right)-\frac{\alpha}{4}\frac{G(\phi_1,\phi_2)\chi^4}{f'^2(\chi,\phi_1,\phi_2)}\right].
	\end{multline}
    Note that we have not employed a new notation to distinguish between metric, and corresponding curvatures, before and after transformation for typographical ease. The second term in (\ref{einmetric}) clearly contains the kinetic term for field $\chi$ and so we can go ahead with the perturbative analysis without having to first find a corresponding constraint equation. Also, instead of expanding the action outright, we'd prefer to only look at the relevant terms. To this end, we perform the perturbative analysis term by term. It is clear that the first term is simply the Einstein-Hilbert action for which the perturbative expansion up to second order is given by,
	\begin{equation}\label{ehpert}
	    \mathcal{L}_1=M_P(\partial_\mu\partial_\nu h^{\mu\nu}-\Box h)-\frac{1}{2}(\partial_\rho h^{\mu\nu}\partial^\rho h_{\mu\nu} -\partial_\mu h\partial^\mu h+2\partial_\mu h \partial_\nu h^{\mu\nu} -2\partial^\mu h^{\nu\rho}\partial_\nu h_{\mu\rho}),
	\end{equation}
	where we have directly substituted for $\kappa$. The second term in this expression gives us the graviton propagator when we add a gauge fixing term to the action ($h=\eta^{\mu\nu}h_{\mu\nu}$). It is very well established that when working on-shell all contributions from gauge fixing terms cancel out eventually. Therefore, we can see directly the following as the graviton propagator:
	\begin{equation}\label{gprop}
	    \left<h^{\rho\sigma}h_{\mu\nu}\right>=\frac{1}{2}\frac{(\delta^\rho_\mu\delta^\sigma_\nu+\delta^\sigma_\mu\delta^\rho_\nu)-\eta^{\rho\sigma}\eta_{\mu\nu}}{k^2} + \mathcal{O}(\lambda),
	\end{equation}
	where $\lambda$ is the gauge parameter introduced to the propagator by the gauge fixing term. The perturbative expansion of the remaining terms yields an unwieldy number of terms, owing to there being three scalars involved in the expansion. Listing them all would be a hassle for both the reader and the author, and so we omit the expansion and detail the process we use.
	
	First, we define a set of dimensionless quantities, inspired by \cite{antoniadis2022addendum},
	\begin{equation}
	    \Bar{G}=1+\frac{\xi\Bar{\phi}_1^2}{M_P^2},\qquad\qquad x^2=\frac{\xi\Bar{\phi}_1^2}{M_P^2 \Bar{G}},
	\end{equation}
	and also introduce the following field transformations,
	\begin{equation}\label{fieldtrans}
	    d\rho=\frac{\sqrt{6}\alpha}{M_P}\frac{\chi d\chi}{\left(1+\frac{\alpha}{M_P^2}\chi^2\right)}, \qquad \psi_1'=\sqrt{\frac{1+6\xi x^2}{\Bar{G}}}\varphi_1,\qquad \psi_2'=\sqrt{\frac{1+6\xi x^2}{\Bar{G}}}\phi_2.
	\end{equation}
	Both these steps help simplify the expressions to a great extent. Listing only the terms second order in dynamical fields from the expansion, we have:
	\begin{align}\label{secondorder}
	    \mathcal{L}^{(2)}=&-\frac{1}{2}\partial_\mu\psi_1'\partial^\mu\psi_1'-\frac{1}{2}\partial_\mu\psi_2'\partial^\mu\psi_2'-\left(\frac{6\xi x^2}{1+6\xi x^2}\right)\partial_\mu\psi_1'\partial^\mu\psi_2'\nonumber\\&
	    -\frac{1}{2}\partial_\mu\rho\partial^\mu\rho-\left(\frac{M_P^2}{2\alpha\Bar{G}^2}\right)\rho^2-\sqrt{\frac{6\xi x^2}{1+6\xi x^2}}\partial_\mu\rho\partial^\mu\psi_1'-\sqrt{\frac{6\xi x^2}{1+6\xi x^2}}\partial_\mu\rho\partial^\mu\psi_2'.
	\end{align}
	Please note that we haven't treated the auxiliary scalar perturbatively. Also, note that compared to the Jordan frame expansion, we shall have to consider far more types of terms that contain vertices that one would not need when computing the required amplitude. The reason for this choice are a few peculiar features to note in (\ref{secondorder}). First, the transformed auxiliary scalar $\rho$ is a massive field and this mass blows up as we try to take the limit $|\alpha|\to 0$, i.e. return to a scalar coupled Einstein-Hilbert action. This is essentially the manifestation of the mass gap problem that plagues higher derivative gravity theories (ex. Starobinsky's model), wherein due to the higher derivative terms, the graviton is no longer a massless propagating field. The scalar $\rho$ retains the issue in this system post the Weyl transformation as well. We shall refer to our inability to take the limit $|\alpha|\to0$ as the `mass-gap' issue for the rest of the paper. Second, note that there is kinetic mixing between the two dynamical scalars and the auxiliary scalar present in (\ref{secondorder}).
    
    Traditionally, this kind of kinetic mixing is dealt with by diagonalizing the kinetic matrix followed by transformations to redefine the theory in terms of canonical fields. Due to the presence of the mass term, when we naively perform this analysis, there appears a mass-type cross-term of the form $c\psi\rho$ (where $c$ represents a constant coefficient). To avoid this kind of an unphysical outcome, we instead try to diagonalize the kinetic and mass matrix $K$, obtained in a similar fashion as the kinetic matrix except with the inclusion of mass terms as well.
    For further typographical ease, we shall denote,
    \begin{align}
        b=\sqrt{\frac{6\xi x^2}{1+6\xi x^2}}, \qquad \text{and} \qquad m^2=\frac{M_P^2}{\alpha\Bar{G}}.
    \end{align}
    For (\ref{secondorder}), we can write the $K$ matrix as,
    \begin{equation}\label{kinmatrix}
    K=\frac{1}{2}
        \begin{pmatrix}
            \Box-m^2 & b\Box & b\Box \\ b\Box & \Box & b^2\Box \\ b\Box & b^2\Box & \Box
        \end{pmatrix}.
    \end{equation}
    We can easily find the eigenvalues as follows:
    \begin{equation}\label{eigen}
        \lambda=\frac{\Box}{2}(1-b^2),\quad\frac{(b^2+2)\Box-m^2\pm\sqrt{b^4\Box^2+8b^2\Box^2+2b^2m^2\Box+m^4}}{4}.
    \end{equation}
    Further, we can find the field transformations that result in the aforementioned eigenvalues using the corresponding eigenvectors. Finding eigenvalues, at least in a way that makes sense from a derivative and tensor notation perspective, is not possible. To do so, we decide to work in different regimes. The simplification of eigenvalues is based on the relationship between the quantities $m^2$ and $b^2\Box$. 
    
    To reiterate, we have no inherent restrictions on the ranges of $\xi$ and $\alpha$. We, however, impose that $\xi>0$. This is because the reference scale for the scalar background is taken to be $\frac{M_P}{\sqrt{\xi}}$ which would turn out to be imaginary for $\xi<0$. The analysis from now on would involve us working in certain ranges of $|\alpha|$ and $\xi$ under large or small background conditions. 
    
    \subsection{$|\alpha|\ll1$}
    
    \subsubsection{Large background}\label{largeback}
    \begin{equation}\label{largebackcond}
        \Bar{\phi}\gg\frac{M_P}{\sqrt{\xi}}\implies\Bar{G}\sim\frac{\xi\Bar{\phi}^2}{M_P^2}\gg 1\implies x^2\to 1\implies b\to \sqrt{\frac{6\xi}{1+6\xi}}.
    \end{equation}
    Instead of using $\xi$ directly, we instead use $b^2$ to define the three classes of results. Here, $b^2\to1$ corresponds to $\xi\gg1$, $b^2\to0$ corresponds to $\xi\to0$ limit, and we additionally include $b^2\to\frac{1}{2}$, which incidentally corresponds to the known conformal limit $\xi\sim\frac{1}{6}$.
    \begin{enumerate}
        \item \textbf{$b^2\to1$}\label{ellmp}
    \end{enumerate}
    In this limit, given that $|\alpha|\ll 1$, it is straightforward to see that $m^2\gg \Box$ and the corresponding eigenvalues can be simplified to,
    \begin{equation}\label{nodiag}
        \lambda=\frac{1}{12\xi}\Box,\ \left(\frac{3}{4}\Box-\frac{m^2}{2}\right) ,\ \frac{3}{4}\Box \ \implies \ \text{two massless + one massive d.o.f.}
    \end{equation}
    And, the corresponding field transformations are:
    \begin{align}\label{transform1}
        \rho&\to\frac{1}{\sqrt{6}}(\psi_2-\psi_1),\nonumber\\
        \varphi_1&\to\frac{\sqrt{\Bar{G}}}{4}\rho',\nonumber\\
        \phi_2&\to-\frac{\sqrt{\Bar{G}}}{4}\rho'+\frac{1}{6}\sqrt{\frac{\Bar{G}}{\xi}}(\psi_1+\psi_2).
    \end{align}
    We choose $\psi_1$ to be massive, while $\rho'$ and $\psi_2$ become massless. Again omitting the transformed Lagrangian, we simply list the vertex factors that may be relevant to the $\psi_1\psi_2\to\psi_1\psi_2$ process that we're focusing on in this work:
    \begin{align}\label{vertfac}
        V_{\psi_1^3}=&-\frac{2i}{9M_P}(k_1\cdot k_2 +k_2\cdot k_3+k_3\cdot k_1)-\frac{iM_P}{18\alpha\Bar{G}^2},\nonumber\\
        V_{\psi_1^2\psi_2}=&\frac{i}{18M_P}(k_{\psi_11}\cdot k_{\psi_2}+k_{\psi_12}\cdot k_{\psi_2})+\frac{2i}{9M_P}(k_{\psi_11}\cdot k_{\psi_12})-\frac{iM_P}{18\alpha\Bar{G}^2},\nonumber\\
        V_{\psi_1\psi_2^2}=&\frac{i}{18M_P}(k_{\psi_1}\cdot k_{\psi_21}+k_{\psi_1}\cdot k_{\psi_22})-\frac{i}{18M_P}(k_{\psi_21}\cdot k_{\psi_22})-\frac{11iM_P}{54\alpha\Bar{G}^2},\nonumber\\
        V_{\psi_2^3}=&-\frac{i}{9M_P}(k_1\cdot k_2+k_2\cdot k_3+k_1\cdot k_3)+\frac{5iM_P}{18\alpha\Bar{G}^2},\nonumber\\
        V_{\psi_1^2\rho'}=&-\frac{\sqrt{\xi}}{3M_P}(k_{\psi_11}\cdot k_{\psi_12}),\nonumber\\
        V_{\psi_1\psi_2\rho'}=&\frac{\sqrt{\xi}}{3M_P}(k_{\psi_1}\cdot k_{\psi_2}),\nonumber\\
        V_{\psi_2^2\rho'}=&-\frac{\sqrt{\xi}}{3M_P}(k_{\psi_21}\cdot k_{\psi_22}),\nonumber\\
        V_{\psi_1^2h_{\mu\nu}}=&-\frac{iM_P^2}{36\alpha\Bar{G}^2}\eta^{\mu\nu},\nonumber\\
        V_{\psi_1\psi_2h_{\mu\nu}}=&\frac{iM_P^2}{36\alpha\Bar{G}^2}\eta^{\mu\nu},\nonumber\\
        V_{\psi_2^2h_{\mu\nu}}=&\frac{i}{3M_P}(k_1^\mu k_2^\nu+k_1^\nu k_2^\mu-\eta^{\mu\nu}k_1\cdot k_2)-\frac{iM_P^2}{36\alpha\Bar{G}^2}\eta^{\mu\nu},\nonumber\\
        V_{\psi_1^2\psi_2^2}=&\frac{8i}{27M_P^2}(k_{\psi_21}\cdot k_{\psi_22})-\frac{17i}{54\alpha\Bar{G}^2}.
    \end{align}
    Note here that all momenta are directed towards the vertex. Also noteworthy is the fact that in the $\xi\gg1$ limit, the third and fourth terms in (\ref{einmetric}) are too small to contribute and can be effectively neglected. Looking at the vertex factors, it is straightforward to conclude that the diagrams that contribute to the $\psi_1\psi_2\to\psi_1\psi_2$ process are as follows: $\psi_1$-exchange $s$, $t$, and $u$ channel diagrams; $\psi_2$-exchange $s$, $t$, and $u$ channel diagrams; $\rho'$-exchange $s$, $t$, and $u$ channel diagrams; graviton-exchange $s$, $t$, and $u$ channel diagrams; and the 4-vertex. Instead of calculating all these diagrams, we shall only look at the types of terms we obtain. 
    
    There are, however, a few key things to note here. The `mass' $m^2=\frac{M_P^2}{\alpha\Bar{G}^2}$ is one gained due to `scalarization' of the higher derivative gravity term. It is straightforward to see that even with the conditions imposed on $|\alpha|$ and $\Bar{G}$, we cannot claim that the usual high-energy approximations can apply here. Considering the case where $k_1$ is attributed to the massive scalar $\psi_1$ and $k_2$ to the massless scalar $\psi_2$, it is easy to see that for an on-shell calculation, $k_1\cdot (k_1+k_2)=k_1^2+k_1\cdot k_2=m^2+\frac{s}{2}$ because $E^2\ll m^2$ up until the Planck scale. Keeping this in mind, we find that the final scattering amplitude for the aforementioned process looks as follows:
    \begin{equation}\label{expr}
        i\mathcal{M}= (A+B\xi)\frac{1}{\alpha\Bar{G}^2} + C\frac{E_E^2}{M_P^2} + D\frac{\alpha\Bar{G}^2E_E^4}{M_P^4} + F\frac{\alpha^2\Bar{G}^2E_E^6}{M_P^6} +\mathcal{O}\left(\frac{E_E^8}{M_P^8}\right),
    \end{equation}
    where $A,\ B,\ C,\ D,$ and $F$ are imaginary dimensionless constants independent of any coupling parameters, and shall be used similarly throughout the rest of the analysis. We have ignored the angular dependence in the final amplitude since we're currently working far from the collinear limit and we're only concerned with violations arising from energy scaling behaviour. We have also ignored some $\mathcal{O}(E_E^{-1})$ terms that exhibited IR violations since that limit isn't under review in this work. 
    
    Now, to go to the Jordan frame as mentioned in Section \ref{unitarity}, we need utilize the field transformation relations for $\varphi_1$ and $\phi_2$ (\ref{transform1}) (we don't use the transformation relation for $\rho$ for obvious reasons). For clarity, the energy relations obtained using both are as follows:
    \begin{align}\label{JEtrans1}
        \varphi_1\to\frac{\sqrt{\Bar{G}}}{4}\rho'&\implies E_J\sim\frac{\sqrt{\xi}\Bar{\phi}_1}{M_P}E_E,\nonumber\\
        \phi_2\to-\frac{\sqrt{\Bar{G}}}{4}\rho'+\frac{1}{6}\sqrt{\frac{\Bar{G}}{\xi}}(\psi_1+\psi_2)&\implies E_J\sim\frac{\Bar{\phi}_1}{M_P}(\sqrt{\xi}+1)E_E\approx E_J\sim\frac{\sqrt{\xi}\Bar{\phi}_1}{M_P}E_E.
    \end{align}
    Thus, the scattering amplitude in the Jordan frame becomes:
    \begin{equation}\label{exprein}
        i\mathcal{M}_J= B\frac{M_P^4}{\alpha\xi\Bar{\phi}_1^4} + C\frac{E_J^2}{\xi\Bar{\phi}_1^2} + D\frac{\alpha E_J^4}{M_P^4} + F\frac{\alpha^2\xi\Bar{\phi}_1^2E_J^6}{M_P^8}+\mathcal{O}(E_J^8),
    \end{equation}
    where we have ignored $A$ since we're working with $\xi\gg1$ in the current case. Using the matching procedure mentioned in Section \ref{unitarity}, taking the limit $\Bar{\phi}_1\to E_J$, it is clear from (\ref{exprein}) that the amplitude is not unitarity violating. As we approach higher energies, the higher order $E$ terms take precedence and simply using the conditions used on the various parameters and backgrounds, we can reach the aforementioned conclusion. This behaviour, however, hinges on the fact that $|\alpha|$ cannot be arbitrarily small. As we take $|\alpha|\to0$, the constant term in (\ref{exprein}) blows up.
    
    \begin{enumerate}
        \setcounter{enumi}{1}
        \item \textbf{$b^2\to\frac{1}{2}$}\label{onebytwo}
    \end{enumerate}
    The matching procedure prescribed in \cite{McDonald_2021} is only valid for $\xi>1$. As soon as we go to $0<\xi<1$, the scalar background becomes trans-Planckian. This limit doesn't cause any major changes in the final expression obtained in (\ref{expr}), even though the full action (\ref{einmetric}) contributes in this case. This is simply because the relevant changes only take place in how the external scalars transform in (\ref{fieldtrans}). For this case, the canonicalization goes as follows:
    \begin{equation}
        \psi_1'=\sqrt{\frac{2}{\Bar{G}}}\varphi_1,\qquad \psi_2'=\sqrt{\frac{2}{\Bar{G}}}\phi_2.
    \end{equation}
    The rest of the procedure remains the same. Thus, (\ref{expr}) becomes:
    \begin{equation}\label{exprmp1}
        i\mathcal{M}_J= A\frac{M_P^4}{\alpha\xi^2\Bar{\phi}_1^4} + B\frac{M_P^4}{\alpha\xi\Bar{\phi}_1^4} + C\frac{E_J^2}{\xi\Bar{\phi}_1^2} + D\frac{\alpha E_J^4}{M_P^4} + F\frac{\alpha^2\xi\Bar{\phi}_1^2E_J^6}{M_P^8}+\mathcal{O}(E_J^8).
    \end{equation}
    In this limit as well, using the same arguments as earlier, we can see that the amplitude (\ref{exprmp1}) doesn't diverge. Additionally, the third term, for which we can no longer take $\Bar{\phi}_1\to E_J$ since the two ranges don't overlap, is well-behaved since $\xi\Bar{\phi}_1\gg M_P^2$, and therefore the theory is safe until the Planck scale. We, however, still see the mass-gap issue from the $|\alpha|\to0$ limit present here.
    \begin{enumerate}
        \setcounter{enumi}{2}
        \item \textbf{$b^2\to0$}\label{esimmp}
    \end{enumerate}
    $b\to0$ basically implies that all off-diagonal terms in the kinetic matrix (\ref{kinmatrix}) disappear in this limit. We, therefore, require no field transformations to deal with the kinetic mixing between the fields and can work with the canonicalized fields in (\ref{fieldtrans}) directly. Also, it is easy to note from perturbative expansion of the Lagrangian (\ref{einmetric}) that the second term is negligible in this scenario. If not for the last term, this case would there be the same as what we expect to find when using the Palatini formalism (as we shall see in an upcoming section). The vertices relevant to the present calculation are as follows:
    \begin{align}
        V_{\psi_1^3}&=\frac{2\sqrt{\xi}}{M_P}(k_1\cdot k_2+k_2\cdot k_3+k_3\cdot k_1),\nonumber\\
        V_{\psi_1^2\psi_2}&=\frac{2\sqrt{\xi}}{M_P}(k_{\psi_11}\cdot k_{\psi_12}),\nonumber\\
        V_{\psi_1\psi_2^2}&=\frac{2\sqrt{\xi}}{M_P}(k_{\psi_21}\cdot k_{\psi_22}),\nonumber\\
        V_{\psi_2^3}&=\frac{2\sqrt{\xi}}{M_P}(k_1\cdot k_2+k_2\cdot k_3+k_3\cdot k_1),\nonumber\\
        V_{\psi_1^2\rho}&=\frac{1}{\sqrt{6}M_P}(k_{\psi_11}\cdot k_{\psi_12}),\nonumber\\
        V_{\psi_2^2\rho}&=\frac{1}{\sqrt{6}M_P}(k_{\psi_21}\cdot k_{\psi_22}),\nonumber\\
        V_{\psi_1^2\psi_2^2}&=\frac{2\xi}{M_P^2}(k_{\psi_11}\cdot k_{\psi_12}+k_{\psi_21}\cdot k_{\psi_22}),\nonumber\\
        V_{\psi_1^2h_{\mu\nu}}&=\frac{1}{2M_P}(k_{\psi_11}^\mu k_{\psi_12}^\nu+k_{\psi_11}^\nu k_{\psi_12}^\mu-\eta^{\mu\nu}k_{\psi_11}\cdot k_{\psi_12}),\nonumber\\
        V_{\psi_2^2h_{\mu\nu}}&=\frac{1}{2M_P}(k_{\psi_21}^\mu k_{\psi_22}^\nu+k_{\psi_21}^\nu k_{\psi_22}^\mu-\eta^{\mu\nu}k_{\psi_21}\cdot k_{\psi_22}).
    \end{align}
    The contributing diagrams in this process are: $\psi_1$-exchange $s$, $t$, $u$ channel diagrams; $\psi_2$-exchange $s$, $t$, $u$ channel diagrams; $\rho$-exchange $t$ channel diagram; and a graviton-exchange $t$ channel diagram.
    Since there are no constant terms in the vertices and the only massive field if $\rho$, we expect no $(\alpha\Bar{G}^2)^{-1}$-type terms in the final expression. The scattering amplitude for such a case looks like:
    \begin{equation}\label{expr1}
        i\mathcal{M}= A\frac{E_E^2}{M_P^2}+B\frac{\alpha\Bar{G}^2E_E^4}{M_P^4} + C\frac{\alpha^2\Bar{G}^4E_E^6}{M_P^6}+\mathcal{O}\left(\frac{E_E^8}{M_P^8}\right).
    \end{equation}
    Using the canonicalization transformations as reference, we can conclude that the Jordan frame version of the amplitude is,
    \begin{equation}\label{expr2}
        i\mathcal{M}= A\frac{E_J^2}{\xi\Bar{\phi}_1^2}+B\frac{\alpha E_J^4}{M_P^4} + C\frac{\alpha^2\xi\Bar{\phi}_1^2E_J^6}{M_P^8}+\mathcal{O}\left(\frac{E_J^8}{M_P^8}\right).
    \end{equation}
    Remembering that in this scenario, matching $\Bar{\phi}_1\to E_J$ isn't possible since their ranges have no overlap, we conclude from the given conditions ($|\alpha|\ll1$, $\xi\Bar{\phi}_1^2\gg M_P^2$) that the scattering amplitude (\ref{expr2}) is well-behaved up until $E_J\to M_P$ provided $\frac{\alpha^2\xi\Bar{\phi}_1^2}{M_P^2}\leq1$.
    
    It is noteworthy that in this scenario, we can take the limit $|\alpha|\to0$ which vanishes all but the first term in (\ref{expr2}). This is simply because the presence of mass in the $\rho$-exchange diagram is responsible for all $\mathcal{O}(E^4)$ terms in (\ref{expr1}). Setting $|\alpha|\to0\implies m_\rho\to\infty$. Due to this the contribution of the $\rho$-exchange diagram becomes zero and only the graviton-exchange diagram remains.

    \subsubsection{Small Background}\label{smallback}
    This limit approximates the electroweak vacuum. Here,
    \begin{equation}\label{smallbackcond}
        \Bar{\phi}_1\ll\frac{M_P}{\sqrt{\xi}}\implies\Bar{G}\sim 1\implies x^2\to \frac{\xi\Bar{\phi}_1^2}{M_P^2}\ll 1\implies \frac{\xi^2\Bar{\phi}_1^2}{M_P^2}\sim 1.
    \end{equation}
    Here, as we shall see, we have to use $\xi$ instead of $b^2$.
    \begin{enumerate}
        \item $\xi\gg1$\label{ellmps}
    \end{enumerate}
    For such a case, we see that
    \begin{equation}
        b\to b'= \sqrt{\frac{\frac{6\xi^2\Bar{\phi}_1^2}{M_P^2}}{1+\frac{6\xi^2\Bar{\phi}_1^2}{M_P^2}}},
    \end{equation}
    since $\left(\frac{6\xi^2\Bar{\phi}_1^2}{M_P^2}\right)$ is of the order $\mathcal{O}(10^0)$. In this regime, $\Bar{\phi}_1$ can assume any value between 0 and $\frac{M_P}{\sqrt{\xi}}<M_P$. All eigenvalues, field transformations and consequently all the vertex factors retain the same form as those listed in (\ref{nodiag}), (\ref{transform1}), and (\ref{vertfac}), respectively (save for modified coefficients). (\ref{expr}) becomes:
    \begin{equation}\label{smallexpr}
        i\mathcal{M}= A\frac{1}{\alpha} + B\frac{E_E^2}{M_P^2} + C\frac{\alpha E_E^4}{M_P^4} + D\frac{\alpha^2E_E^6}{M_P^6} +\mathcal{O}\left(\frac{E_E^8}{M_P^8}\right).
    \end{equation}
    Going back to the Jordan frame in this scenario is straightforward. The field transformations remain the same as in (\ref{transform1}), but since $\Bar{G}\to1$, we have $E_J=E_E$. i.e. (\ref{smallexpr}) is the final expression. As we can observe, the expression is safe from any scaling violations, but due to the constant term $\alpha^{-1}$, unitarity is violated nonetheless at all energies. This effectively prohibits $\xi\gg1$ as a viable parameter range. Also, $|\alpha|\to0$ is impossible here too.
    
    \begin{enumerate}
        \setcounter{enumi}{1}
        \item $0<\xi<1$\label{xi01}
    \end{enumerate}
    Here, we can safely take $b\to0$ and effectively ignore all the off-diagonal terms in the kinetic matrix, implying that there's again no need for field transformations. Also, since $\Bar{G}\to1$, we don't even need canonicalization of fields. We can thus directly use the perturbative expansion without having to resort to the matching procedure. The range of the scalar field background up until $\frac{M_P}{\sqrt{\xi}}>M_P$.
    
    Here again, we need to define two distinct limits of $\xi$: $\xi\to\frac{1}{6}$ and $\xi\to0$. In the latter case, we get the same vertices and scattering amplitude as Section \ref{esimmp}. For the former case, we also have terms corresponding to vertices relevant for $s$ and $u$ channel $\rho$-exchange diagrams. Nevertheless, the scattering amplitude for both these cases remains the same and can be expressed, in the Jordan frame, as:
    \begin{equation}\label{expr3}
        i\mathcal{M}= A\frac{E_J^2}{M_P^2}+B\frac{\alpha E_J^4}{M_P^4} + C\frac{\alpha^2E_J^6}{M_P^6}+\mathcal{O}\left(\frac{E_J^8}{M_P^8}\right).
    \end{equation}
    The amplitude is clearly well-behaved as we take $E_J\to M_P$. Note that in case, we can safely take the $|\alpha|\to0$ limit without any issues, similar to Section \ref{esimmp}.
    
    \subsection{$|\alpha|\to1$}
    \subsubsection{Large Background}\label{largeback1}
    Proceeding the same way as earlier, we again have three limits of $b^2$.
    \begin{enumerate}
        \item $b^2\to1,\frac{1}{2}$
    \end{enumerate}
    In this limit, $m^2$ is no longer too large. Now, as $E\ll M_P$, we find that $m^2\sim b^2\Box$, meaning obtaining meaningful eigenvalues from (\ref{eigen}) is not feasible. This, however, changes as $E\to M_P$, where $m^2\ll b^2\Box$. But as we approach the Planck scale, it is known that loop corrections are no longer perturbative and can dictate the way that the amplitude behaves. So, calculating the tree-level amplitude around that scale would simply be an exercise and not reveal any meaningful results. Omitting this limit entirely, we move on to the next case.
    
    \begin{enumerate}
        \setcounter{enumi}{1}
        \item $b^2\to0$
    \end{enumerate}
    We can neglect the off-diagonal terms in $K$ and perform the analysis directly, obtaining similar scattering amplitude as in (\ref{expr2}), though the conclusions are wildly different. The amplitude diverges in the $|\alpha|\to1$ limit, implying unitarity violations as $E\to M_P$.
    
    \subsubsection{Small Background}\label{smallback1}
    \begin{enumerate}
        \item $\xi\gg1$
    \end{enumerate}
    For $E^2\ll M_P$, we have $m^2\sim b^2\Box$, meaning the eigenvalues are not properly defined. However, as we get to $E\to M_P$, we find that we again recover $m^2\gg b^2\Box$, same as the analysis covered in Section \ref{smallback} for the same limit. While the amplitude retains the same form as (\ref{smallexpr}), it narrowly remains unitary since $|\alpha|\to1$.
    \begin{enumerate}
        \setcounter{enumi}{1}
        \item $0<\xi<1$
    \end{enumerate}
    Here, too, the resulting amplitude is the same as that obtained in the corresponding limit in Section \ref{smallback} and it remains unitary as well.

    \subsection{$|\alpha|\gg1$}
    \subsubsection{Large Background}\label{largeback2}
    \begin{enumerate}
        \item $b^2\to1$
    \end{enumerate}
    In this limit, we find that $m^2\ll b^2\Box$. Thus, the eigenvalues (\ref{eigen}) simplify to give:
    \begin{equation}\label{b21}
        \lambda=\frac{1}{12\xi}\Box, \left(\frac{3}{2}\Box-\frac{m^2}{4}\right), \frac{-m^2}{4} \implies \text{one massless+one massive+one non-dynamical d.o.f.}
    \end{equation}
    It is clear from the eigenvalues that one of the components of the singlet scalar turns completely non-dynamical in this limit, implying that the process $\psi_1\psi_2\to\psi_1\psi_2$ is no longer possible. 
    \begin{enumerate}
        \setcounter{enumi}{1}
        \item $b^2\to\frac{1}{2}$
    \end{enumerate}
    The arguments posed above apply here as well. The eigenvalues, however, are different:
    \begin{equation}\label{b212}
        \lambda=\frac{1}{4}\Box, \left(\frac{9}{8}\Box-\frac{m^2}{4}\right), \left(\frac{1}{8}\Box-\frac{m^2}{4}\right) \implies \text{one massless + two massive d.o.f.}
    \end{equation}
    The final amplitude calculated using these has a form similar to (\ref{exprmp1}). However, since $|\alpha|\gg1$, the amplitude diverges far below the Planck scale.
    
    \begin{enumerate}
        \setcounter{enumi}{2}
        \item $b^2\to0$
    \end{enumerate}
    Again, no diagonalization of $K$ is necessary in this limit, and we arrive at the same amplitude as (\ref{expr2}). Due to the current limit on $\alpha$, the amplitude diverges in this limit.

    \subsubsection{Small Background}\label{smallback2}
    \begin{enumerate}
        \item $\xi\gg1$
    \end{enumerate}
    For $E\to M_P$, we again find that $m^2\sim b^2\Box$, meaning the eigenvalues are poorly defined. Below that scale, $m^2\gg b^2\Box$ and predictions can still be made in the same way as in Section \ref{smallback}. Going $E\to M_P$ using this expression immediately results in unitarity violations, thus enforcing the fact that straightforward extrapolation of the amplitude is not possible and that information in that regime is missing. 
    \begin{enumerate}
        \setcounter{enumi}{1}
        \item $0<\xi<1$
    \end{enumerate}
    The amplitude here is the same as that obtained in (\ref{expr3}), though it violates unitarity as we approach the Planck regime.

    \section{Palatini Formulation}\label{pala}
	Now moving to Palatini formalism with a torsion-free spacetime, following the Weyl transformation, (\ref{lagrange}) becomes:
	\begin{equation}\label{einpala}
	    S=\int d^4x \sqrt{-g} \left[\frac{1}{2}M_P^2R-\frac{1}{2}\left(\frac{\partial_\mu\varphi_1\partial^\mu\varphi_1}{f'(\chi,\phi_1,\phi_2)}\right)-\frac{1}{2}\left(\frac{\partial_\mu\phi_2\partial^\mu\phi_2}{f'(\chi,\phi_1,\phi_2)}\right)-\frac{W(\chi,\phi_1,\phi_2)}{f'^2(\chi,\phi_1,\phi_2)}\right].
	\end{equation}
	Varying this action w.r.t. $\chi^2$, we obtain the following constraint equation:
	\begin{equation}\label{constrainteq}
	    f'(\chi,\phi_1,\phi_2)=\frac{M_P^4G(\phi_1,\phi_2)}{M_P^4-\alpha(\partial_\mu\varphi_1\partial^\mu\varphi_1+\partial_\mu\phi_2\partial^\mu\phi_2)}.
	\end{equation}
	For more details about this, refer to \cite{Das2021}. Using this, we can go ahead and eliminate $\chi$ from the action and rewrite it as:
	\begin{multline}\label{palatiniaction}
	    S=\int d^4x \sqrt{-g}\left[\frac{1}{2}M_P^2R-\frac{1}{2G(\phi_1,\phi_2)}(\partial_\mu\varphi_1\partial^\mu\varphi_1+\partial_\mu\phi_2\partial^\mu\phi_2)\right.\\\left.+\frac{\alpha}{4M_P^2G(\phi_1,\phi_2)}(\partial_\mu\varphi_1\partial^\mu\varphi_1+\partial_\mu\phi_2\partial^\mu\phi_2)^2\right].
	\end{multline}
	At this point, an inquisitive reader may question why we chose to eliminate $\chi$. The reason for doing that is because of the absence of a kinetic term for field $\chi$. Due to that, even though there is some $\alpha$ dependence in the second term of the action, it is of no actual use to us at the present stage since it is always accompanied by the non-propagating $\chi$.
	
	We can now go ahead and expand the action perturbatively and find the relevant propagators and vertices. The first term is again the Einstein-Hilbert action that we have already expanded in (\ref{ehpert}) and found the corresponding propagator for in (\ref{gprop}).
	Similarly, doing the perturbative expansion for the remaining terms of the action (\ref{einpala}),
	\begin{align}\label{kinpert}
	    \mathcal{L}_2=&\sqrt{-g}\left[-\frac{1}{2G(\phi_1,\phi_2)}(\partial_\mu\varphi_1\partial^\mu\varphi_1+\partial_\mu\phi_2\partial^\mu\phi_2)\right]\nonumber\\
	    =&-\frac{1}{2\Bar{G}}\partial_\mu\varphi_1\partial^\mu\varphi_1-\frac{1}{2\Bar{G}}\partial_\mu\phi_2\partial^\mu\phi_2+\frac{\sqrt{\xi}x}{M_P\Bar{G}^{3/2}}\varphi_1\partial_\mu\varphi_1\partial^\nu\varphi_1+\frac{\sqrt{\xi}x}{M_P\Bar{G}^{3/2}}\phi_2\partial_\mu\varphi_1\partial^\nu\varphi_1\nonumber\\
	    &+\frac{\sqrt{\xi}x}{M_P\Bar{G}^{3/2}}\varphi_1\partial_\mu\phi_2\partial^\nu\phi_2+\frac{\sqrt{\xi}x}{M_P\Bar{G}^{3/2}}\phi_2\partial_\mu\phi_2\partial^\nu\phi_2+\frac{\kappa}{2\Bar{G}}(h^{\mu\nu}-\frac{1}{2}\eta^{\mu\nu}h)\partial_\mu\varphi_1\partial_\nu\varphi_1\nonumber\\
	    &+\frac{\kappa}{2\Bar{G}}(h^{\mu\nu}-\frac{1}{2}\eta^{\mu\nu}h)\partial_\mu\phi_2\partial_\nu\phi_2+\frac{\xi}{2M_P^2\Bar{G}^2}\left(1-\frac{3x^2}{\Bar{G}}\right)\varphi_1^2\partial_\mu\phi_2\partial^\mu\phi_2\nonumber\\
	    &+\frac{\xi}{2M_P^2\Bar{G}^2}\left(1-\frac{3x^2}{\Bar{G}}\right)\phi_2^2\partial_\mu\varphi_1\partial^\mu\varphi_1,\\
	    \mathcal{L}_3=&\sqrt{-g}\left[\frac{\alpha}{4M_P^2G(\phi_1,\phi_2)}(\partial_\mu\varphi_1\partial^\mu\varphi_1+\partial_\mu\phi_2\partial^\mu\phi_2)^2\right]=\frac{\alpha}{2M_P^4\Bar{G}}\partial_\mu\varphi_1\partial^\mu\varphi_1\partial_\nu\phi_2\partial^\nu\phi_2.
	\end{align}
	We restate the fact that we are only listing the terms that would be relevant to us for tree-level scattering amplitude computations. Now, the kinetic terms of scalar fields can be brought to their canonical form using the following transformations:
	\begin{equation}\label{canon1}
	    \varphi_1\to\sqrt{\Bar{G}}\phi_1',\qquad\phi_2\to\sqrt{\Bar{G}}\phi_2',
	\end{equation}
	which consequently redefines the corresponding vertices involving the scalar field, and also provides us with a relation between the Jordan and Einstein frame energies, same as the one obtained in (\ref{JEtrans1}). Note that all information about the Starobinsky term and its coupling with the scalars is present in $\mathcal{L}_3$, which, for the process under review, represents a correction to the 4-vertex. The calculation from this point seems pretty straightforward. The vertex factors in momentum space then become:
	\begin{align}
	    V_{\phi_1'^3}=&\frac{2i\sqrt{\xi}x}{M_P}(k_1\cdot k_2+k_2\cdot k_3+k_3\cdot k_1),\nonumber\\
	    V_{\phi_1'^2\phi_2'}=&\frac{2i\sqrt{\xi}x}{M_P}(k_{\phi_1'1}\cdot k_{\phi_1'2}),\nonumber\\
	    V_{\phi_1'\phi_2'^2}=&\frac{2i\sqrt{\xi}x}{M_P}(k_{\phi_2'1}\cdot k_{\phi_2'2}),\nonumber\\
	    V_{\phi_2'^3}=&\frac{2i\sqrt{\xi}x}{M_P}(k_1\cdot k_2+k_2\cdot k_3+k_3\cdot k_1),\nonumber\\
	    V_{\phi_1'^2\phi_2'^2}=&\frac{2i\xi}{M_P^2}\left(1-\frac{3x^2}{\Bar{G}}\right)(k_{\phi_1'1}\cdot k_{\phi_1'2}+k_{\phi_2'1}\cdot k_{\phi_2'2})+\frac{2i\alpha\Bar{G}}{M_P^4}(k_{\phi_1'1}\cdot k_{\phi_1'2})(k_{\phi_2'1}\cdot k_{\phi_2'2}),\nonumber\\
	    V_{\phi_1'^2h_{\mu\nu}}=&\frac{i}{2M_P}(k_1^\mu k_2^\nu+k_1^\nu k_2^\mu-\eta^{\mu\nu}k_1\cdot k_2),\nonumber\\
	    V_{\phi'^2h_{\mu\nu}}=&\frac{i}{2M_P}(k_1^\mu k_2^\nu+k_1^\nu k_2^\mu-\eta^{\mu\nu}k_1\cdot k_2),
	\end{align}
	where all momenta are directed towards the vertex. Now that we have the vertices, we can calculate the scattering amplitude for the process $\phi_1'\phi_2'\to\phi_1'\phi_2'$. It is straightforward to see that the tree-level diagrams that contribute to this process are: $\phi_1'^2\phi_2'^2$-vertex, $\phi_1'$-exchange, $\phi_2'$-exchange in $s$, $t$, and $u$ channels, and $h_{\mu\nu}$-exchange $t$ channel diagrams . We see again that since the scalar field is massless, the $s$, $t$, $u$ channel diagrams involving $\phi_1'$- and $\phi_2'$-exchange cancel out (owing to the identity: $s+t+u=\Sigma m_{ext}=0$). The relevant information is present only in the 4-vertex and graviton-exchange diagrams.
	The scattering amplitude has the form:
	\begin{equation}\label{expr4}
        i\mathcal{M}= (A+B\xi)\frac{E_E^2}{M_P^2}+C\frac{\alpha\Bar{G}E_E^4}{M_P^4}.
    \end{equation}
    Now, we shall work out the various cases we encountered in the metric formulation case. The $|\alpha|\to0$ limit is safe throughout the Palatini formulation for this action, as per both the Lagrangian (\ref{palatiniaction}) and scattering amplitude (\ref{expr4}). 
    
    \subsection{$|\alpha|\ll1$}
    
	\subsubsection{Large Background}
	Conditions here remain the same as (\ref{largebackcond}). In this limit, (\ref{expr4}) becomes:
	\begin{equation}\label{expr5}
        i\mathcal{M}= (A+B\xi)\frac{E_J^2}{\xi\Bar{\phi}_1^2}+C\frac{\alpha E_J^4}{\xi\Bar{\phi}_1^2M_P^2}.
    \end{equation}
    \begin{enumerate}
        \item $\xi\gg 1$
    \end{enumerate}
	In this case, the background has a lower bound at $\frac{M_P}{\sqrt{\xi}}<M_P$ and the matching procedure ($\Bar{\phi}_1\to E_J$) suggested in \cite{McDonald_2021} is possible. It is straightforward to see that the amplitude is well-behaved in this scenario.
	\begin{enumerate}
	    \setcounter{enumi}{1}
	    \item $0<\xi<1$
	\end{enumerate}
	Here, the scalar background becomes trans-Planckian with lower bound at $\frac{M_P}{\sqrt{\xi}}>M_P$ and \cite{McDonald_2021}'s procedure is no longer valid. It is, however, still easy to see that since $\xi\Bar{\phi}_1^2\gg M_P^2$, the amplitude is again well-behaved up to the Planck scale.

    \subsubsection{Small Background}\label{smallbackpala}
    Conditions here remain the same as (\ref{smallbackcond}). Here, the amplitude remains the same as (\ref{expr4}) because in this limit $E_J=E_E$. As in the metric case, the matching procedure is rendered useless here. The scattering amplitude is given as:
    \begin{equation}\label{expr6}
        i\mathcal{M}= B\frac{\xi E_J^2}{M_P^2} + C\frac{\alpha E_J^4}{M_P^4}.
    \end{equation}
    Working out the various limits of $\xi$ as before,
    \begin{enumerate}
	    \item $\xi\gg 1$
	\end{enumerate}
	Looking at this limit empirically, we can effectively ignore the first term and conclude that the second term violates unitarity as $E\to\frac{M_P}{\sqrt{\xi}}$, while the third term is safe up to the Planck scale.
	
	\begin{enumerate}
	    \setcounter{enumi}{1}
	    \item $0<\xi<1$
	\end{enumerate}
    It is clear from (\ref{expr6}) that the amplitude is safe up to the Planck scale.
	
	Broadly speaking, the amplitudes and observations about the applicability of McDonald's matching procedure remain the same in the other limits of $|\alpha|$ as well.

	\subsection{$|\alpha|\to1$}
	
	\subsubsection{Large Background}
	Using similar arguments as posed for the $|\alpha|\ll1$ case, we can easily see that amplitudes retain unitarity throughout the full range of $\xi$.
	
	\subsubsection{Small Background}
	We find that the scattering amplitudes violate unitarity for $\xi\gg1$ as we approach the Planck scale due to the first term in (\ref{expr6}). This violation is avoided for $0<\xi<1$ limit.
	
	\subsection{$|\alpha|\gg1$}
	
	\subsubsection{Large Background}
	This regime is a little non-trivial in comparison. Here, we find that the unitarity violations from the scattering amplitude in (\ref{expr5}) can be avoided in the limit $\xi\gg1$ if $|\alpha|\leq\xi$. This is because matching is applicable in this limit. Now, for the limit $0<\xi<1$, since matching is no longer valid, we find that the correct constraint to preserve unitarity is $\frac{|\alpha| M_P^2}{\xi\Bar{\phi}_1^2}\leq1$. 
	
	\subsubsection{Small Background}
	It is straightforward to see from (\ref{expr6}) that unitarity is violated for all $\xi$ for this case.

    \section{Application to Inflationary Scenarios}
    As mentioned earlier, we didn't assume a particular form for the potential in (\ref{action}) so that the parameters remain flexible in their range. Now, we can impose restrictions based on known inflation scenarios.
    \subsection{Matthew et al.'s Inflation Scenario}\label{matt}
    First, we refer to \cite{Mathew:2017lvh} for this section, specifically, the case where there is no $\phi^4$ term in the potential (mentioned as a special case in \cite{Mathew:2017lvh}). For such a scenario, they conclude that $\alpha$ must assume a small, negative value in order to ensure a safe exit from inflation, while no constraints were imposed on $\xi$. We again refer to the Tables (\ref{metrictable}) and (\ref{palatable}) for information on these limits.

    It is clear to see from the tables that for metric formulation, unitarity is preserved in both inflation and reheating epochs for $0<\xi<1$, subject to some constraints mentioned both in Section \ref{largeback} and Table (\ref{metrictable}) as $\xi\ll1$. The limit $\xi\to\frac{1}{6}$, however, is free from any such uncertainties and the corresponding amplitude is unitary throughout. But, considering the safety of taking $|\alpha|\to0$ limit, and effectively bypassing the mass-gap issue for higher derivative gravity, we find that the viable limit for $\xi$ is only $\xi\ll1$, which is subjected to the constraints mentioned earlier.
    
    Similarly, for Palatini formulation, the favorable limits are found to be $|\alpha|\ll1$ and $0<\xi<1$. Additionally, taking the limit $|\alpha|\to0$ is safe for Palatini throughout.
    
    \subsection{Higgs'-like Inflation Scenario}\label{higg}
    Next, looking at Higgs' inflation scenario, for which the potential is of the form:
    \begin{equation}
        V(|\Phi|)=\lambda\left((\Phi^\dagger\Phi)^2-\frac{v^2}{2}\right),
    \end{equation}
    where we have identified $\Phi$ as the Higgs' complex singlet scalar, and $v$ represents the vacuum expectation value of the Higgs' field. Traditionally, inflation models work in the unitary gauge, such that the Higgs' complex singlet simplifies to a real scalar $\phi_1$ driving inflation. A similar case has been analysed in \cite{SBudhi:2019vln}, where the authors found that using values $v=246$ GeV $(\ll M_P)$ and $\lambda\simeq 0.13$ (in accordance with observed values $v$ and Higgs' mass $= \lambda v^2\simeq125$ GeV), the coupling parameter $\xi$ would assume values $>\mathcal{O}(10^4)$, similar to Higgs' inflation scenario.

    We have listed briefly the results from this paper in the form of Tables (\ref{metrictable}) and (\ref{palatable}). For this particular case, we need only look at the case where $\xi\gg1$ or equivalently $b^2\to1$. In metric formulation, we see that for $|\alpha|\ll1$ the theory violates unitarity as we transition from inflation to reheating stage. For the other two limits of $|\alpha|$, the present analysis was insufficient to make any claims on the preservation of unitarity. Similarly, for Palatini formulation, we see that the transition is unsafe throughout as the small background limit violates unitarity for the entire range of $|\alpha|$. We do, however, see that taking the limit $|\alpha|\to0$ is safe in the Palatini formulation, while it is unsafe for the metric case.

    \section{Discussion and Conclusions} \label{result}
    In this work, we have calculated the tree-level scattering amplitude for a 2-scalar $\to$ 2-scalar process for a scalar-Starobinsky inflation model with a modification involving a nonminimal coupling between the $R^2$ term and the scalar field (\ref{action}). We use the arguments put forward by \cite{Hertzberg:2010dc,McDonald_2021,Antoniadis2021,antoniadis2022addendum} to list the regimes where the process would not violate unitarity in the UV limit and accordingly narrow down the constraints on the coupling parameters in the action. We have performed the calculations assuming no prior constraints on $\xi$ and $\alpha$, other than the positivity of $\xi$, which was explained in Section \ref{metric}. We follow the methodology of \cite{antoniadis2022addendum} and additionally work out the case where the scalar field background is $\ll \frac{M_P}{\sqrt{\xi}}$. The results for metric and Palatini formulations have been summarized neatly in Tables (\ref{metrictable}) and (\ref{palatable}), respectively.
    
    \begin{table}
        \centering
        \begin{tabular}{ |c|c|c|c|c| }
        \hline
        \multicolumn{3}{|c|}{} & Unitarity up to $M_P$ & $|\alpha|\to0$ limit \\
        \hline
        \multirow{5}{4em}{$|\alpha|\ll1$} & \multirow{3}{8em}{Large Background} & $b^2\to1$ & Safe & Unsafe \\
                                                                              \cline{3-5} 
                                        &                                     & $b^2\to1/2$ & Safe & Unsafe \\
                                                                              \cline{3-5} 
                                        &                                     & $b^2\to0$ & Safe if $\frac{\alpha^2\xi\Bar{\phi}_1^2}{M_P^2}\leq1$ & Safe \\
                                        \cline{2-5}
                                        & \multirow{2}{8em}{Small Background} & $\xi\gg1$ & Unsafe & Unsafe \\
                                                                              \cline{3-5}
                                        &                                     & $0<\xi<1$ & Safe & Safe \\
        \hline
        \multirow{4}{4em}{$|\alpha|\to1$} & \multirow{2}{8em}{Large Background} & $b^2\to1,1/2$ & NA \\
                                                                              \cline{3-4}
                                        &                                     & $b^2\to0$ & Unsafe \\
                                        \cline{2-4}
                                        & \multirow{2}{8em}{Small Background} & $\xi\gg1$ & Safe \\
                                                                              \cline{3-4}
                                        &                                     & $0<\xi<1$ & Safe \\
        \cline{1-4}
        \multirow{5}{4em}{$|\alpha|\gg1$} & \multirow{3}{8em}{Large Background} & $b^2\to1$ & NA \\
                                                                              \cline{3-4}
                                        &                                     & $b^2\to1/2$ & Unsafe \\
                                                                              \cline{3-4}
                                        &                                     & $b^2\to0$ & Unsafe \\
                                        \cline{2-4}
                                        & \multirow{2}{8em}{Small Background} & $\xi\gg1$ & NA \\
                                                                              \cline{3-4}
                                        &                                     & $0<\xi<1$ & Unsafe \\
        \cline{1-4}
        \end{tabular}
        \caption{Metric Formulation: Results for whether the theory is unitary for the given conditions. Also includes results for whether it is safe to take the $|\alpha|\to0$ limit.}
        \label{metrictable}
    \end{table}
    
    \begin{table}
        \centering
        \begin{tabular}{ |c|c|c|c|c| }
        \hline 
        \multicolumn{3}{|c|}{} & Unitarity up to $M_P$ & $|\alpha|\to0$ limit \\
        \hline
        \multirow{5}{4em}{$|\alpha|\ll1$} & \multirow{2}{8em}{Large Background} & $\xi\gg1$ & Safe & Safe \\
                                                                              \cline{3-5}
                                        &                                     & $0<\xi<1$ & Safe & Safe \\
                                        \cline{2-5}
                                        & \multirow{2}{8em}{Small Background} & $\xi\gg1$ & Unsafe & Safe \\
                                                                              \cline{3-5}
                                        &                                     & $0<\xi<1$ & Safe & Safe \\
        \hline
        \multirow{4}{4em}{$|\alpha|\to1$} & \multirow{2}{8em}{Large Background} & $\xi\gg1$ & Safe \\
                                                                              \cline{3-4}
                                        &                                     & $0<\xi<1$ & Safe \\
                                        \cline{2-4}
                                        & \multirow{2}{8em}{Small Background} & $\xi\gg1$ & Unsafe \\
                                                                              \cline{3-4}
                                        &                                     & $0<\xi<1$ & Safe \\
        \cline{1-4}
        \multirow{4}{4em}{$|\alpha|\gg1$} & \multirow{2}{8em}{Large Background} & $\xi\gg1$ & Safe if $|\alpha|\leq\xi$ \\
                                                                              \cline{3-4}
                                        &                                     & $0<\xi<1$ & Safe if $\frac{|\alpha| M_P^2}{\xi\Bar{\phi}_1^2}\leq1$ \\
                                        \cline{2-4}
                                        & \multirow{2}{8em}{Small Background} & $\xi\gg1$ & Unsafe \\
                                                                              \cline{3-4}
                                        &                                     & $0<\xi<1$ & Unsafe \\
        \cline{1-4}
        \end{tabular}
        \caption{Palatini Formulation: Results for whether the theory is unitary for the given conditions. Also includes results for whether it is safe to take the $|\alpha|\to0$ limit.}
        \label{palatable}
    \end{table}
    
    The inflation era is characterized by $\Bar{\phi}_1\sim\frac{M_P}{\sqrt{\xi}}$ which then immediately switched to $\Bar{\phi}_1\to0$ as soon as it ends and reheating begins. The analysis and arguments put forward in \cite{Hertzberg:2010dc} were for the reheating era, also referred to as the electroweak vacuum in \cite{Antoniadis2021}. While using a truncated perturbative expansion, we need to differentiate between the limits $E\ll\Bar{\phi}_1$ (which is equivalent to $\left<\phi\right>\ll\Bar{\phi}_1$) and $E\gg\Bar{\phi}_1$ (which is equivalent to $\left<\phi\right>\gg\Bar{\phi}_1$). Since we're working with an infinite series sum, however, both these limits would be equivalent at the perturbative expansion level, and we need to look elsewhere to enforce the condition $\Bar{\phi}\to0$ condition.
    
    This is where the small background limits for various cases in Sections \ref{metric} and \ref{pala} become useful. In these sections, we use $\Bar{\phi}_1\ll\frac{M_P}{\sqrt{\xi}}$ from which we can easily take the limit $\Bar{\phi}_1\to0$ irrespective of the range of $\xi$. We also analysed from the forms of the scattering amplitudes whether taking the limit $|\alpha|\to0$ was viable in each case. This is necessary because we need to be able to recover Einstein-Hilbert gravity safely in physical processes, due to the incredible observational accuracy exhibited by the theory. There are screening mechanisms that ensure that such `mass-gap' issues don't show up at low-energy scales such as Vainshtein screening, which was elaborated in our previous work \cite{Panda:2021cbf}. In sections \ref{matt} and \ref{higg}, we form relevant conclusions without considering such screening processes.
    
    There are some ranges of $|\alpha|$ and $\xi$ for which predictions couldn't be made in the metric formulation (marked NA in the Table (\ref{metrictable})) due to computational restraints stemming from our inability to find proper eigenvalues to the kinetic matrix $K$. For $|\alpha|\to1$, this range is from $\xi\sim\frac{1}{6}$ to $\xi\to\infty$ in the large background limit. Similarly for $|\alpha|\gg1$, the corresponding range is $\xi\gg1$ in both the large and small background limits. These limits could just as well be safe until the Planckian regime, but we can't make any definitive comments at the current stage. We leave the unitarity analysis of these limits as a future work.
    

    \section*{Acknowledgement}
	We would like to thank the authors of \cite{Antoniadis2021,antoniadis2022addendum} I. Antoniadis, A. Guillen, and K. Tamvakis for their valued comments and suggestions on our work. A.V. would also like to thank G. S. Punia for helpful discussions. Some part of the calculations in this paper have been carried out in MATHEMATICA using the xAct packages: xTensor \cite{martin-garcia} and xPert \cite{Brizuela:2008ra}. This work is partially supported by DST (Govt. of India) Grant No. SERB/PHY/2021057.
	
	\bibliographystyle{unsrtnat}
	\bibliography{main}
\end{document}